\documentclass[11pt]{article}
\usepackage{graphicx}

\newsavebox{\hflrar}
\sbox{\hflrar}{\makebox[0pt][l] {${\scriptstyle
\leftharpoonup}$}{${\scriptstyle \rightharpoonup}$}}

\def \to {\rightarrow}

\begin{document}

\begin{center}
{\Large\bf Reply to ``the Comment on `Gauge Invariance and $k_T$-Factorization of Exclusive Processes'"
} \vskip 10mm
F. Feng$^1$, J.P. Ma$^1$ and Q. Wang$^{2}$    \\
{\small {\it $^1$ Institute of Theoretical Physics, Academia Sinica,
Beijing 100190, China }} \\
{\small {\it $^2$ Department of Physics and Institute of Theoretical
Physics, Nanjing Normal University, Nanjing, Jiangsu 210097, P.R.China}} \\
\end{center}
\vskip 5mm

\begin{abstract}
A new method is proposed to calculate wave functions in
$k_T$-factorization in \cite{LiMi} as a comment about our paper
\cite{FMW}. We point out that the results obtained with the
method are in conflict with the translation invariance and depend on
the chosen contours for loop-integrals. Therefore, the method is in
principle unacceptable and the results with the method cannot be correct.
% PACS numbers:
\end{abstract}
\vskip 0.5cm
\par
Recently we have pointed out in \cite{FMW} that the $k_T$-factorization performed with off-shell
partons is in general gauge variant and violated, because the perturbative
coefficient function $H$ contains gauge-dependent light-cone singularities
and gauge-dependent $\mu$-dependence.
The $k_T$-factorization for the transition $\pi^0 +\gamma^* \to \gamma$ has been examined in \cite{NLi}
in Feynman gauge at one-loop level.
It is claimed in \cite{NLi} that the perturbative
coefficient function $H$ calculated with off-shell partons is gauge-invariant.
A proof of the gauge invariance at any-loop level is given in \cite{NLi}, where one shows
the gauge invariance at $N+1$-loop level with the induction method by assuming
the gauge invariance at $N$-loop level.
But, the gauge invariance of the one-loop $H^{(1)}$ is not shown.
This motivates us to study the problem in \cite{FMW}.
\par
To determine
the perturbative coefficient function $H$ in the $k_T$-factorization one replaces
$\pi^0$ with an off-shell quark pair. With the quark pair one calculates the form factor and
the wave function. $H$ receives contributions from the form factor and the wave function.
In \cite{FMW} it is shown with covariance arguments that the form factor does not contain
the gauge-dependent light-cone singularities.
The singularities come from wave functions.
In the updated comment made by Li and Mishima\cite{LiMi}
a new method is proposed to calculate wave functions with off-shell partons. It is found
that the gauge-dependent light-cone singularities disappear. In this updated reply we point
out unacceptable flaws of the method.
\par
The definition of the wave function for $\pi^0$ moving in the $z$-direction is \cite{FMW,NLi}:
\begin{eqnarray}
\phi(x, k_T,\zeta, \mu) = \ \int \frac{ d z^- d^2 z_\perp}{(2\pi )^3}  e^{i x P^+ z^- - i \vec z_\perp\cdot \vec
k_T}
\langle 0 \vert \bar q(0) L_u^\dagger (\infty, 0)
  \gamma^+ \gamma_5 L_u (\infty,z) q(z) \vert \pi^0(P) \rangle\vert_{z^+=0} .
\end{eqnarray}
We have taken the light-cone coordinate system in which $\pi^0$ has the large momentum component
$P^+$. $L_u$ is a gauge link along the direction $u^\mu =(u^+,u^-,0,0)$
with $u^-\ll u^+$ and $\zeta$ is defined as $\zeta^2 \approx
4 (u\cdot P)^2/ u^2$.
In the definition there is the momentum conservation
in the $+$- and transverse direction due to the translation invariance.
Because of this and $z^+ =0$ all $-$-components
of momenta in loops are integrated over from $-\infty$ to $\infty$.
It is noted that $\phi$ only depends on $x, k_T,\zeta$ and $\mu$ as indicated in the
above.
\par
To extract the one-loop coefficient $H^{(1)}$ one needs to calculate
$\phi$ of an off-shell parton pair instead of $\pi^0$ at one-loop, as
required in the $k_T$-factorization\cite{NLi}.  Then one calculates
the convolution of $\phi$ with $H^{(0)}$, i.e., $\phi^{(1)}\otimes
H^{(0)}$ as a part of contributions to $H^{(1)}$. The mentioned
singularity in $\phi$ comes from Fig.2a, 2b and 2c in
\cite{FMW,LiMi}, where one has an integral of $q^-$ from $-\infty$
to $\infty$. $q^-$ is the component of the momentum $q$ carried by
the exchanged gluon in the figures, the component $q^+$ and
$q_\perp$ are fixed in $\phi$ by the momentum conservation. The
integral can be performed by a standard contour integral in the
$q^-$ complex-plane. One can takes a closed contour like those
indicated in Fig.5 of \cite{LiMi} for $-R\leq q^- \leq R$ at first
step by adding semicircles. After the integration one takes the
limit $R\to \infty$. In this limit one finds that all contributions
to $\phi$ from semicircles in Fig.5 of \cite{LiMi} are zero. The
remaining contributions to $\phi$ are only from poles in the complex
plane and contain the light-cone singularity\cite{FMW,LiMi}. In the
limit the contribution $\phi_{2b}$ from Fig.2b in \cite{FMW,LiMi} to $\phi$ is nonzero
only for $0<q^+ <k_1^+$  with $k_1^\mu =(k_1^+,0,\vec k_{1\perp})$
as the momentum carried by the off-shell quark.
\par
In \cite{LiMi} it is suggested that one should keep these
contributions to $\phi$ from the semicircles to calculate the
convolution $\phi^{(1)}\otimes H^{(0)}$, which contains integrations over $x$ and
$k_T$. $x$ is related to $q^+$ through $x P^+ = k_1^+ -q^+$ in
\cite{FMW,LiMi}. After finishing the integrations in the convolution
one then takes the limit $R\to \infty$. In this way the light-cone
singularity disappears because the contributions from semicircles
given in Fig.5 \cite{LiMi} give nonzero contributions to the
convolution in the limit. Since the limit is taken after the
convolution, this implies that  $\phi$ depends on an extra parameter
$R$ which is in fact a cut-off for the large $\vert q^- \vert $. The
existence of a finite $R$ actually is in conflict with the
translational invariance. We take the contribution $\phi_{2b}$ as an
example to illustrate this in more detail in the following.
\par
It is well-known that the wave function in Eq.(1) is zero for $x<0$
and $x>1$. The boundary $x=1$ can be obtained by inserting a
complete sum of intermediate states in Eq.(1) and using the
translational invariance. With
the finite $R$ $\phi_{2b}$ is not zero for $q^+<0$. This nonzero
contribution is crucial to cancel the light-cone
singularity of $\phi_{2b}$ with $q^+ >0$\cite{LiMi}.
If one takes the limit $R\to \infty$ before
the convolution, $\phi_{2b}$ is zero for $q^+<0$.
From the translation invariance one can
show that $\phi_{2b}$ or contributions from similar class of
diagrams are zero for $q^+ <0$ without any calculation.
The contributions from Fig.2b in \cite{FMW,LiMi} and
its higher-order corrections, in which there is no gluon
attached to the antiquark line and the gauge link $L_u(\infty,z)$,
are actually proportional to the contributions to a distribution of
a single quark with the momentum $k_1$.  The distribution can be
defined by replacing $\pi^0$ in Eq.(1) with the single
off-shell quark and delete some operators in Eq.(1) as in the
following:
\begin{eqnarray}
 \int \frac{ d z^- d^2 z_\perp}{(2\pi )^3} e^{i y k_1^+ z^- - i \vec
z_\perp\cdot \vec k_T} \langle 0 \vert  \left ( L_u^\dagger (\infty, 0) -1 \right )
   q(z) \vert q (k_1) \rangle\vert_{z^+=0} .
\end{eqnarray}
$\phi_{2b}$ is proportional to one of the leading order contributions of the above distribution
with $x P^+ = y k_1^+$.
Inserting a complete sum of intermediate states and using the
translation invariance, we have that the  above distribution
must be zero for $ y >1$. Here $y$ is related to $q^+$ through $ y k_1^+ = k_1^+ -q^+$.
One can conclude without any calculation
that the above distribution and $\phi_{2b}$
must be zero for $q^+
<0$, or the contributions from similar diagrams are zero if the out
going quark line carries a $+$-momentum component is larger than
that of the incoming quark line. This conclusion is in agreement
with the physical expectation that partons entering hard scattering
have positive $+$-energy. Therefore, $\phi_{2b}\neq 0$ with $q^+ <0$
is in conflict with the translation invariance. To avoid the conflict
one has to take the limit $R\to\infty$ before the convolution. This results
in $\phi_{2b} = 0$ for $q^+ <0$.
\par
In fact there is certain ambiguity
in $\phi_{2b}$ if one does not take the limit
$R\to\infty$.
In the method, one actually changes the order of the integrations in
the convolution $\phi_{2b}\otimes
H^{(0)}$ with the $q^-$-integration in $\phi_{2b}$
by implementing certain prescriptions, where one first deforms
the contours along the real axis in the $q^-$-plane for the
$q^-$-integral in $\phi$ to those in Fig.5 of \cite{LiMi}.
With these contours one
finds the nonexistence of the light-cone singularity. If one flips
the closed contour in Fig.5b of \cite{LiMi}, i.e., by changing the
contour in the upper-half  $q^-$-plane into the lower-half
$q^-$-plane, one will find a different result from Fig.5b than that
given in Eq.(12) of \cite{LiMi} by noting that fact: The
contribution from the single pole in the upper-half  $q^-$-plane is
the same as that from the double pole in the lower-half $q^-$-plane,
and with the flipped contour the corresponding integral of $\theta$
from $\pi \to 0$ in Eq.(12) of \cite{LiMi} is changed to that of
$\theta$ from $-\pi \to 0$. The final results with the flipped contour
will still contain the light-cone singularity.
The cancelation of the light-cone singularity only happens with certain contours
like those given in Fig.5 of \cite{LiMi}. This is in conflict with the general expectation
that physical results should not depend contours used for loop-integrals.
\par
It is interesting to note that one will not
find the singularity in the convolution if one simply changes the
order of the integrations of  $\phi_{2b}\otimes
H^{(0)}$ by ignoring the constraint of allowed
$q^+$-regions from the translational invariance, where one first integrates over $q^+$
by taking the region with $q^+<0$ into account
and then perform the integration over $q^-$. However, from the translational invariance
we know that the region
with $q^+<0$ should be excluded in the integration of $q^+$. Taking this into account,
$\phi_{2b}$ and its convolution with $H^{(0)}$ has the mentioned light-cone singularity.
\par
The above discussion tells that the
final results with the method will depend on chosen contours. The
method implies a modification for the wave function of parton states
through implementing a cut-off and specially chosen contours. It is
unclear how to implement the modification to the wave function at hadron
level, i.e., to that in Eq.(1). With the implement the calculated wave function is in
fact no longer the wave function employed in the
$k_T$-factorization in Eq.(1). Since the results obtained with the method are
in conflict with the translational invariance and depend on the
chosen contours, the proposed method is in principle unacceptable.
\par
To conclude: In the suggested method of \cite{LiMi} an extra cut-off
is introduced into wave functions and special contours for
$-$-components of loop momenta are taken. The results obtained with
the method are in conflict with the translation invariance and
depend on the chosen contours. Therefore, the method is in principle
unacceptable and the obtained result with the method can not be correct.

\par\vskip20pt
\noindent
{\bf\large Acknowledgments}

\par
We thank Prof. M. Yu for useful discussions.
This work is supported by National Nature Science Foundation of P.R. China (No. 10721063, 10575126, 10747140).
\par

%%%%%%%%%%%%%%%%%%%%%%%%%%%%%%%%%%%%%%%%%%%%%%%%%%%%%%%%%%%%%%%%%%%%%%%%%%%%
%%%%%%%%%%%%%%%%%%%%%%%%%%%%%%%%%%%%%%%%%%%%%%%%%%%%%%%%%%%%%%%%%%%%%%%%%%%%
%%%%%%%%%%%%%%%%%%%%%%%%%%%%%%%%%%%%%%%%%%%%%%%%%%%%%%%%%%%%%%%%%%%%%%%%%%%%

\end{document}